# Analyse multigraduelle OLAP

Gilles Hubert, Olivier Teste

Université de Toulouse – IRIT (UMR 5505)
118, Route de Narbonne – 31062 Toulouse cedex 9 (France)
{Gilles.Hubert, Olivier.Teste}@irit.fr

**Résumé.** Les systèmes décisionnels reposent sur des bases de données multi-dimensionnelles qui offrent un cadre adéquat aux analyses OLAP. L'article présente un nouvel opérateur OLAP nommé « BLEND » rendant possible des analyses multigraduelles. Il s'agit de transformer la structuration multidimen-sionnelle lors des interrogations pour analyser les mesures selon des niveaux de granularité différents recombinées comme un même paramètre. Nous menons une étude des combinaisons valides de l'opération dans le contexte des hiérarchies strictes. Enfin, une première série d'expérimentations implante l'opération dans le contexte R-OLAP en montrant le faible coût de l'opération.

## 1 Introduction

Les systèmes d'aide à la prise de décision connaissent un essor important en raison de leur capacité à supporter efficacement les analyses sur les données disponibles dans les organisations. Ces systèmes décisionnels sont élaborés à partir du système opérationnel d'une organisation : les données identifiées comme pertinentes pour les décideurs sont extraites, transformées, puis chargées (Vassiliadis, *et al.*, 2002) dans un espace de stockage appelé entrepôt de données (« data warehouse »). Afin d'améliorer l'interrogation et l'analyse de ces données entreposées, des techniques d'organisation des données spécifiques ont été développées (Kimball, 1996) reposant sur des bases de données multidimensionnelles (BDM). Ce type de modélisation considère la donnée à analyser comme un point dans un espace à plusieurs dimensions, formant ainsi un cube de données (Gray, *et al.*, 1996). Les décideurs qui utilisent ces systèmes visualisent un extrait des cubes de données, généralement une tranche à deux dimensions d'un cube. A partir de cette structure, appelée table multidimensionnelle (TM) (Gyssens et Lakshmanan, 1997), le décideur peut interagir au travers d'opérations de manipulation. Les opérations les plus emblématiques sont les forages qui consistent à modifier la graduation d'un axe d'analyse (niveaux de granularité) et les opérations de rotation qui consistent à changer de tranche de cube. On parle d'analyse en ligne ou encore de processus OLAP (« On-Line Analytic Processing ») (Ravat, *et al.*, 2008).

Cet environnement offre un cadre adéquat aux analyses des décideurs, cependant la structure imposée peut s'avérer imparfaite ou devenir obsolète. Considérons des montants de ventes analysés en fonction de clients français et de clients américains. Dans ce cadre, un décideur peut vouloir utiliser la graduation en fonction du pays pour les clients français tandis qu'il souhaite utiliser simultanément une graduation différente, par exemple les états américains pour les clients des États-Unis. En effet, pour certaines analyses, il est nécessaire de comparer un pays tel que la France avec des entités géographiques différentes comme les états américains afin de comparer des informations équivalentes en superficie, en taille de population, etc. Ce type de manipulation est difficile à mettre en œuvre voire impossible avec les systèmes actuels. L'objectif de cet article est de proposer une solution permettant ces manipulations qualifiées de multigraduelles.

## 1.1 État de l'art

Il existe principalement deux approches pour la modélisation des BDM : une approche reposant sur la métaphore du cube de données suivant laquelle la BDM est représentée par des cubes, et une approche dite de modélisation multidimensionnelle où la BDM est décrite par un schéma en étoile ou en constellation (Kimball 1996). Plusieurs travaux de synthèses du domaine (Chaudhuri et Dayal, 1997), (Vassiliadis et Sellis, 1999) et études comparatives (Abelló, *et al*., 2006), (Ravat, *et al*., 2008) sont disponibles.

Lors de manipulation OLAP, un des premiers travaux étend l'opération d'agrégation dans le contexte des manipulations OLAP (Gray, *et al*., 1996). Un grand nombre d'opérations ont été définies, cependant par manque de consensus sur un modèle de référence, les propositions d'opérations OLAP n'ont toujours pas été clairement identifiées et définies au sein d'une algèbre à l'instar de l'approche relationnelle. Une étude comparative des nombreuses propositions existantes est disponible dans (Romero et Abelló, 2007).

A notre connaissance aucune proposition ne permet de répondre à notre problématique. Les solutions les plus proches proposent des mécanismes visant à personnaliser une BDM en transformant ses valeurs et ses structures. Dans (Espil et Vaisman, 2001) le langage à base de règles IRAH est introduit pour permettre aux décideurs de réorganiser les regroupements des valeurs entre deux graduations. Cette approche ne permet pas de transformer les structures hiérarchiques des graduations initialement mises en place dans la BDM. Plus récemment, (Favre, *et al*., 2007) ont introduit un mécanisme à base de règles « Si-Alors » afin d'intégrer des connaissances utilisateurs pour faire évoluer le schéma de la BDM. Ce mécanisme permet de manière personnalisée d'ajouter de nouvelles graduations. Bien que ces solutions permettent une certaine adaptation de la BDM, elles se heurtent à deux problèmes : premièrement, le processus de transformation est délicat et fastidieux car reposant sur des définitions de règles exprimées par le décideur, et deuxièmement, la cohérence et la confiance aux données décisionnelles entreposées ne sont plus garanties. Le fait d'introduire des moyens directs d'accès en modification des valeurs rend inopérant les processus de nettoyage et de consolidation habituels.

D'autres travaux dans le contexte de l'évolution des BDM ont proposé des opérations de transformation des hiérarchies initialement modélisées (Blaschka, *et al.*, 1999) (Hurtado, *et al*., 1999) (Eder, *et al*., 2003). Dans (Blaschka, *et al*., 1999), une opération d'insertion d'un nouveau paramètre est présentée. L'opération « Relate Levels », définie dans (Hurtado, *et al*., 1999), permet de transformer l'organisation des paramètres des hiérarchies. D'autres opérations de transformation (« Split », …) portant sur les valeurs contenues dans les paramètres sont décrites dans (Eder, *et al*., 2003). Ces travaux offrent un cadre permettant l'évolution des hiérarchies, mais ne correspond pas réellement à des transformations multigraduelles. Ces opérations peuvent être détournées pour transformer la BDM. Cependant, notre objectif est différent puisqu'il vise à permettre de réorganiser les regroupements des valeurs entre deux graduations, et ceci, lors du processus d'analyse, sans impacter physiquement les données stockées dans la BDM.

### 1.2 Contributions et organisation

La contribution essentielle de cet article est la proposition d'une nouvelle manipulation dans les BDM facilitant les analyses multigraduelles. Une analyse multigraduelle combine une même mesure analysée en fonction de données issues de plusieurs paramètres : par exemple, nous rendons possibles l'analyse de superficies agricoles en fonction de valeurs géographiques de niveaux différents telles que les superficies des États-Unis et de l'Europe.

Nous étendons l'algèbre OLAP que nous avons définie dans le laboratoire (Ravat, *et al.*, 2008) par l'opérateur d'analyse multigraduelle « BLEND ». Nous effectuons une étude des différents cas possibles de fonctionnement de l'opérateur dans le contexte des hiérarchies strictes (Malinowski et Zimanyi, 2006). Nous proposons un fonctionnement par transformation de la hiérarchie courante et nous établissons les contours de l'opérateur en identifiant ses limites. Enfin, nous expérimentons l'opération dans le contexte de mise en œuvre R-OLAP.

Un avantage de la solution proposée, est de rendre possible ce type d'analyse au moment de la manipulation des données alors qu'elle nécessite dans un contexte classique la réorganisation complète des données et des processus ETL d'alimentation associés. La phase de construction d'une BDM est une tâche lourde difficilement réitérable en fonction de chaque besoin d'analyse. L'application de ces transformations au niveau des analyses sans impacter l'organisation réelle des données, facilite le partage de la BDM.

La section 2 présente le modèle de représentation conceptuelle d'une BDM que nous adoptons. Nous définissons un nouvel opérateur, appelé « BLEND » dans la section 3. Nous montrons les différents cas possibles de manipulations multigraduelles qu'il autorise dans le contexte de hiérarchisation stricte des dimensions (Malinowski et Zimanyi, 2006). La section 4 décrit la mise en œuvre de l'opérateur dans le contexte R-OLAP.

## 2 Modélisation et manipulation OLAP

Cette section introduit les différentes définitions des concepts permettant de représenter les éléments d'une BDM (Ravat, *et al.*, 2008). Notre modèle repose sur une représentation des données par une constellation (Kimball 1996) regroupant un ensemble de faits associés à des dimensions partagées ou spécifiques.

**Définition 1.** Une constellation C est définie par ($N^C$, $F^C$, $D^C$, $Star^C$) où $N^C$ est le nom de la constellation, $F^C = \{F_1,…, F_p\}$ est un ensemble de faits, $D^C = \{D_1,…, D_q\}$ est un ensemble de dimensions, et $Star^C : F^C \rightarrow 2^{D^C}$ [1] est une fonction associant les faits aux dimensions.

**Définition 2.** Une dimension D modélise un axe d'analyse et se définit par ($N^D$, $A^D$, $H^D$, $I^D$) où $N^D$ est le nom de la dimension, $A^D = \{a1,…, a_u\} \cup \{All, Id\}$ est un ensemble d'attributs, $H^D = \{h^D_1,…, h^D_v\}$ est un ensemble de hiérarchies, et $I^D = \{I^D_1, I^D_2,…\}$ est l'ensemble des instances de D. Les hiérarchies organisent les attributs d'une dimension, appelés paramètres, de la graduation la plus fine jusqu'à la graduation la plus générale. Ainsi une hiérarchie définit les chemins de navigation valides sur un axe d'analyse.

**Définition 3.** Une hiérarchie $h^D_i$ est un chemin élémentaire acyclique partant du paramètre racine, noté Id, jusqu'au paramètre extrémité, noté All. Elle est définie par ($N^{hD}_i$, $Param^{hD}_i$, $Suppl^{hD}_i$) où $N^{hD}_i$ est le nom de la hiérarchie, $Param^{hD}_i : P \rightarrow P$ ($P \subseteq A^D$) est une fonction décrivant la hiérarchie des attributs, appelés paramètres de la hiérarchie, $Suppl^{hD}i$ :

---

[1] La notation $2^E$ représente l'ensemble des parties d'un ensemble E.

Analyse multigraduelle OLAP

P$\rightarrow 2^{(AD-P)}$ est une fonction spécifiant les attributs faibles qui complètent la sémantique des paramètres (un ensemble, éventuellement vide, d'attributs faibles est associé à un paramètre).

**Définition 4.** Un fait F représente un ensemble d'indicateurs relatifs à un sujet d'analyse et se définit par le quadruplet ($N^F$, $M^F$, $I^F$, $IStar^F$) où $N^F$ est le nom du fait, $M^F = \{f_1(m_1),\ldots, f_w(m_w)\}$ est un ensemble de mesures (ou indicateurs) agrégées selon une fonction $f_i \in \{$AVG, SUM, MAX, MIN, COUNT,...$\}$, $I^F = \{I^F_1, I^F_2,\ldots\}$ est l'ensemble des instances de F, et $IStar^F$ est une fonction associant les instances de $I^F$ aux instances des dimensions liées au fait.

Adossés au modèle de représentation, nous définissons, d'une part, une structure de visualisation appelée TM qui centre l'analyse sur un fait en fonction de deux dimensions sélectionnées. D'autre part, un opérateur de construction et un ensemble d'opérateurs de manipulation sont disponibles. Nous ne détaillons pas dans cet article ces opérations ; une étude exhaustive est disponible dans (Ravat, *et al.*, 2008).

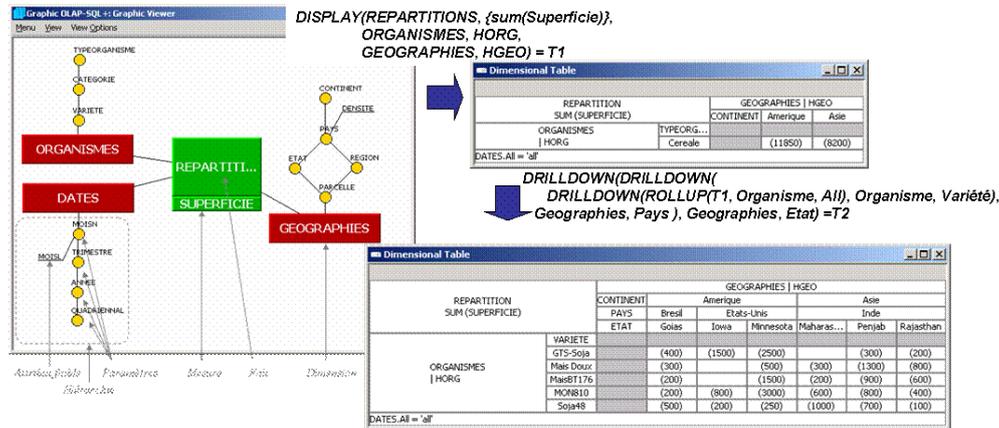

FIG. 1 – *Exemple d'étoile (constellation à un seul fait) et de TM obtenues par manipulations.*

**Exemple 1.** Une BDM supportant des analyses sur la répartition de parcelles OGM dans le monde est définie au travers d'une constellation formée d'un fait et de trois dimensions (*cf.* figure 1) ; nous adoptons un formalisme graphique proche de celui proposé par (Golfarelli, *et al.*, 1998). Un décideur peut alors visualiser le contenu du fait et des dimensions au travers d'une TM ; la table T1 présentée affiche la répartition des superficies par continent et par type d'organisme. Par une combinaison d'opérations, le décideur transforme cette T1 pour visualiser dans T2 les superficies par état des pays en fonction des différentes variétés afin de comparer les variétés OGM (GTS-Soja, Maïs BT176 et Mon 810) aux variétés « classiques ».

## 3 L'opérateur « BLEND »

Afin de répondre à notre problématique d'analyse multigraduelle, nous définissons une opération permettant de transformer les paramètres des dimensions. Cette opération baptisée « BLEND » s'applique sur les TM de sorte à modifier les entêtes des lignes ou des colonnes.

## 3.1 Opérateur algébrique

**Définition 5.** L'opération de transformation multigraduelle d'une TM est définie par :
$$\text{BLEND}(T_{SRC}, D, P_{sup}(s_{sup}), P_{inf}(s_{inf}), \text{pred}) = T_{RES}$$

- $T_{SRC} = (S_{SRC}, L_{SRC}, C_{SRC}, R_{SRC})$ est la TM source à transformer,
- $D \in \{DL_{SRC}, DC_{SRC}\}$ est l'une des dimensions de la TM $T_{SRC}$,
- $P_{sup}$ et $P_{inf}$ sont des paramètres consécutifs affichés de la dimension D tels que $P_{sup}$ est le paramètre hiérarchiquement supérieur à $P_{inf}$,
- $s_{sup} \in \{+,-\}$ et $s_{inf} \in \{+,-\}$ sont des estampilles indiquant la conservation (+) ou non (-) du paramètre associé dans $T_{RES}$ ; l'utilisation des estampilles et leurs différentes combinaisons sont étudiées de manière exhaustive dans la section suivante (3.2),
- pred est un prédicat de sélection permettant de déterminer les valeurs issues des paramètres $P_{sup}$ et $P_{inf}$ pour construire le domaine de définition du nouveau paramètre noté $P_{sup}\_P_{inf}$ dans $T_{RES}$,
- $T_{RES}$ est la TM résultat.

Le prédicat pred sert à calculer les ensembles $E_{sup}$ et $E_{inf}$, qui regroupent les valeurs issues des paramètres $P_{sup}$ et $P_{inf}$ participant à la construction du domaine du nouveau paramètre :
- $E_{sup}$ contient les valeurs de $P_{sup}$ sélectionnées par pred,
- $E_{inf}$ contient les valeurs de $P_{inf}$ sélectionnées par $\neg$pred.

**Contrainte 1.** Le prédicat noté pred dans la définition de l'opérateur « BLEND » est valide si et seulement si $E_{sup} \cap \text{parent}(E_{inf}) = \varnothing$ avec :
- parent($E_{inf}$) désigne les valeurs de dom($P_{sup}$) liées à $E_{inf}$,
- dom($P_{sup}$) désigne le domaine de définition de $P_{sup}$.

Par abus de langage, nous dirons que pred doit définir deux ensembles de valeurs « disjoints » au regard de l'organisation hiérarchique.

**Contrainte 2.** La composition d'opérateurs « BLEND » n'est pas commutative. L'utilisateur doit construire ses manipulations en tenant compte de l'ordre des paramètres $P_{sup}$ et $P_{inf}$, mais également de l'ordre des combinaisons des transformations multigraduelles.

## 3.2 Cas de transformation

L'opérateur « BLEND » modifie la hiérarchie existante en substituant un nouveau paramètre à l'un des paramètres existants (ou les deux) ou en intégrant un nouveau paramètre en plus des paramètres existants. L'intérêt de l'opération réside dans la transformation de la hiérarchie existante remplaçant la hiérarchie initiale considérée obsolète directement dans la TM par l'utilisateur sans imposer une reconstruction de la BDM.

L'intégration du nouveau paramètre peut s'effectuer selon quatre scenarii :
- soit le paramètre remplace les deux existants $P_{sup}$ et $P_{inf}$ (TAB. 1 - a),
- soit le paramètre remplace le paramètre $P_{inf}$ (TAB. 1 - b),
- soit le paramètre remplace le paramètre $P_{sup}$ (TAB. 1 - c),
- soit le paramètre s'intercale entre les paramètres $P_{sup}$ et $P_{inf}$ (TAB. 1 - d).

Les estampilles ajoutées aux deux paramètres $P_{sup}$ et $P_{inf}$ indiquent le scénario choisi. L'estampille (-) indique que le paramètre ne doit pas apparaître dans le résultat tandis que l'estampille (+) indique le contraire. Ceci permet de transformer deux paramètres en créant un nouveau paramètre multigraduel, tout en choisissant de maintenir tout ou partie des possibilités de navigations initiales (avec les opérations de forages). Dans le tableau 1, le scenario

Analyse multigraduelle OLAP

(a) fait disparaître les possibilités de forer sur les pays et les états (seul le paramètre multi-graduel reste disponible) tandis que le scénario (d) maintient les deux paramètres initiaux.

Il est important de noter que nous ne présentons ici que les possibilités qui maintiennent des hiérarchies strictes (Malinowski et Zimanyi, 2006) dans lesquelles toute valeur du paramètre inférieur ne peut être liée qu'à une seule valeur du paramètre supérieur.

TAB. 1 – *Quatre possibilités de modification de la hiérarchie.*

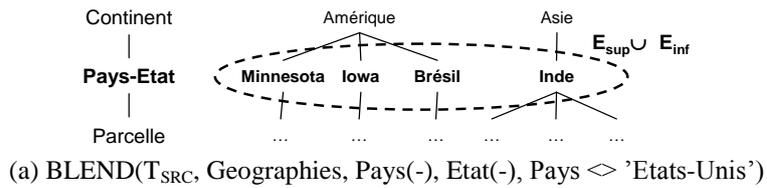

(a) BLEND($T_{SRC}$, Geographies, Pays(-), Etat(-), Pays <> 'Etats-Unis')

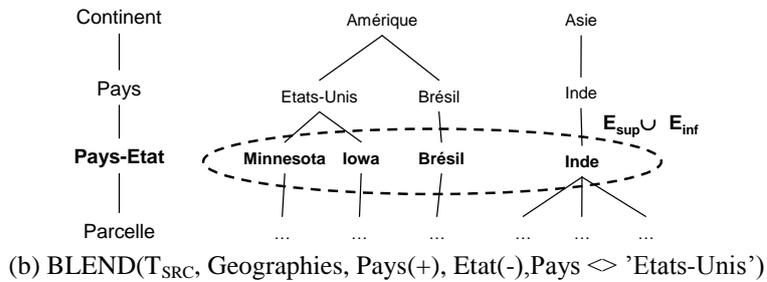

(b) BLEND($T_{SRC}$, Geographies, Pays(+), Etat(-), Pays <> 'Etats-Unis')

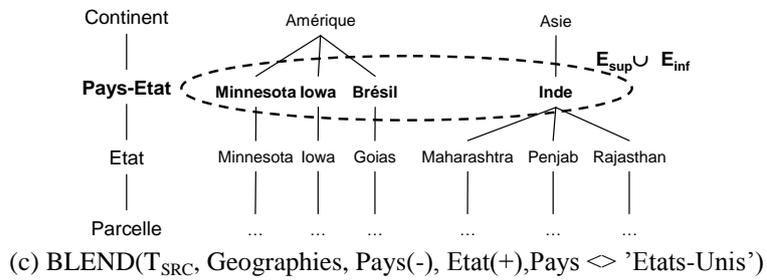

(c) BLEND($T_{SRC}$, Geographies, Pays(-), Etat(+), Pays <> 'Etats-Unis')

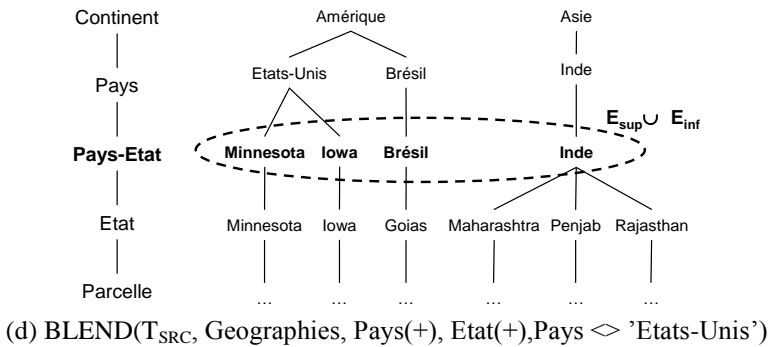

(d) BLEND($T_{SRC}$, Geographies, Pays(+), Etat(+), Pays <> 'Etats-Unis')

### 3.3 Propriété de fermeture de l'opérateur

La définition de l'opérateur « BLEND » respecte la propriété de fermeture : il s'applique sur une TM et produit une nouvelle TM. Cette propriété permet l'enchaînement de plusieurs opérations afin d'opérer des transformations complexes à partir de l'opération élémentaire.

**Exemple 2.** Considérons une analyse complexe dans laquelle un décideur souhaite comparer les superficies de céréales entre les états américains, un pays tel que le Brésil et le continent asiatique. Cette analyse est multigraduelle sur trois niveaux puisqu'elle fait intervenir, un continent, un pays, et des états américains (subdivisions d'un pays). A partir de la TM T2, nous enchaînons les deux opérations de transformation multigraduelle suivantes :

BLEND(BLEND(T2, Geographies, Pays(-), Etat(-), Pays <> 'Etats-Unis'), Geographies, Continent(-), Pays-Etat(-),Continent = 'Asie') = T3

Les figures suivantes illustrent l'enchaînement des deux opérations avec les transformations multigraduelles induites.

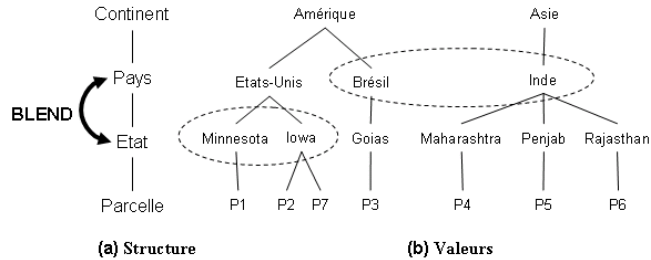

FIG. 2 – *Structuration initiale de Geographies dans T2.*

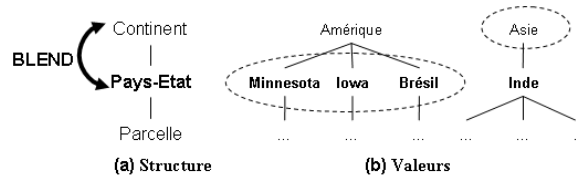

FIG. 3 – *Structuration intermédiaire de Geographies après le premier « BLEND » figure 2.*

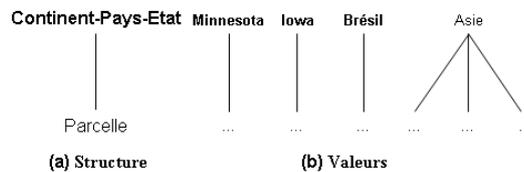

FIG. 4 – *Structuration finale de Geographies dans T3 après le second « BLEND » figure 3.*

L'enchaînement des deux BLEND donne lieu à une transformation multigraduelle des données de T2. La table résultat T3 est présentée à la figure 5.

Analyse multigraduelle OLAP

FIG. 5 – *Principe de transformations multigraduelles.*

## 3.4 Cas particuliers de l'opérateur

### 3.4.1 Sélection vide

L'expression du prédicat pred dans l'opérateur « BLEND » peut conduire à une sélection vide de valeurs sur $P_{sup}$ ou $P_{inf}$. Si l'ensemble $E_{sup}$, respectivement $E_{inf}$, est vide, alors l'opération reste valide. Ce cas particulier consiste à supprimer le paramètre $P_{sup}$, respectivement $P_{inf}$. Notons que ce résultat peut s'obtenir par combinaison d'autres opérateurs de l'algèbre (Ravat, *et al.*, 2008). Par définition, le cas où $E_{sup}$ et $E_{inf}$ seraient simultanément vides ne peut survenir.

**Exemple 3.** Considérons une nouvelle opération « BLEND » qui cherche à combiner les pays ayant une forte densité de population (Densité > 20) avec les états de pays dont la densité est moindre. L'opération suivante est exprimée.

BLEND(T3, Geographies, Pays($s_{sup}$), Etat($s_{inf}$), Densité>20) avec $s_{sup} \in \{+,-\}$ et $s_{inf} \in \{+,-\}$.

Sachant que les États-Unis, le Brésil et l'Inde ont respectivement les densités de 31.15, 21.60 et 300.24 hab/km$^2$, le prédicat Densité > 20 aboutit au calcul des ensembles de valeurs suivants : $E_{sup}$ = {'Etats-Unis', 'Brésil', 'Inde'} et $E_{inf} = \varnothing$. Ce cas particulier où $E_{inf}$ est vide revient à conserver tous les pays tout en faisant disparaître de l'analyse les états associés.

### 3.4.2 Paramètre racine

L'opération « BLEND » appliquée à la racine de la dimension, induit la transformation multigraduelle de la dimension et le recalcul des valeurs des mesures associées. En effet, la suppression des valeurs du paramètre racine nécessite l'agrégation des valeurs des mesures associées. Chaque valeur agrégée est reliée à la valeur du paramètre supérieur.

**Exemple 4.** Considérons l'opération « BLEND » faisant intervenir le paramètre racine Parcelle suivant laquelle un décideur souhaite comparer les superficies entre les états américains et toutes les parcelles d'autres pays. L'opération suivante est exprimée.

BLEND(T3, Geographies, Etat($s_{sup}$), Parcelle($s_{inf}$), Pays = 'Etats-Unis').

Les ensembles $E_{sup}$={'Minnesota', 'Iowa'} et $E_{inf}$={'P3', 'P4', 'P5', 'P6'} sont calculés.

Dans la TM résultat, les valeurs des mesures initialement associées aux parcelles des États-Unis ('P1', 'P2' et 'P7') sont agrégées pour être reliées aux valeurs des états sélectionnés dans $E_{sup}$. La fonction utilisée est celle spécifiée lors de la construction de la TM initiale par le constructeur DISPLAY (figure 1). Dans cet exemple, la fonction SUM est utilisée. Ce principe d'utilisation, lors des manipulations suivantes, de la même fonction d'agrégat que celle utilisée initialement est identique aux opérations de forage. Il peut être envisagé d'étendre l'opérateur pour permettre à l'utilisateur de spécifier une fonction d'agrégation différente de celle utilisée pour construire la table source.

## 4 Expérimentation en environnement R-OLAP

L'opération « BLEND » est implantée au sein de l'outil Graphic-OLAP que nous développons dans notre laboratoire avec le langage Java au dessus du SGBD Oracle. Les principes de modélisation et de manipulation des BDM que nous définissons sont mis en œuvre dans un contexte R-OLAP. L'architecture repose sur un stockage relationnel de données et de métadonnées tout en présentant diverses interfaces à l'utilisateur.

La constellation de faits et de dimensions est implantée sous forme de tables : un ensemble de métatables décrit la structure multidimensionnelle et un ensemble de tables stocke les données décisionnelles disponibles pour l'analyse. Pour simplifier, notre présentation est limitée aux tables de stockage des données détaillées ; nous n'abordons pas la problématique d'optimisation par des vues matérialisées (Zhuge, *et al.*, 1998) (Kotidis et Roussopoulos, 1999). Dans ce cadre simplifié, les interrogations effectuées par l'utilisateur sont traduites par une requête SQL d'extraction sur les tables stockant les données décisionnelles.

**Exemple 5.** Le schéma en étoile de la figure 1 est stocké en R-OLAP comme suit :
DATES(**id_dates**, moisn, moisl, trimestre, annee, quadriennal)
ORGANISMES(**id_organismes**, variete, categorie, typeorganisme)
GEOGRAPHIES(**id_geographies**, parcelle, etat, region, pays, densite, continent)
REPARTITION(**id_repartition**, **id_dates#**, **id_organismes#**, **id_geographies#**, superficie)

Reconsidérons les deux opérations « BLEND » illustrées figures 2, 3 et 4. La TM T3 de la figure 5 est obtenue à partir du résultat de requêtes d'extraction générées par Graphic-OLAP. Le tableau suivant montre les requêtes SQL générées pour chaque opération :

TAB. 2 – *Traduction SQL de l'opération BLEND.*

| BLEND(T2, Geographies, Pays(-), Etat(-), Pays <> 'Etats-Unis') = $T_i$ | BLEND($T_i$, Geographies, Continent(-), Pays-Etat(-), Continent = 'Asie') = T3 |
|---|---|
| Traduction SQL : <br> **SELECT SUM**(superficie) **AS** superficie, <br>  continent, pays **AS** pays_etat, variete <br> **FROM** (**SELECT** superficie, continent, pays, variete <br>   **FROM** T2 **WHERE** pays <> 'Etats-Unis' <br>  **UNION ALL** <br>   **SELECT** superficie, continent, pays, variete <br>   **FROM** T2 **WHERE NOT** (pays <> 'Etats-Unis')) <br> **GROUP BY** continent, etat, variete; | Traduction SQL : <br> **SELECT SUM**(superficie) **AS** superficie, <br>  continent **AS** continent_pays_etat, variete <br> **FROM** (**SELECT** superficie, continent, pays, variete <br>   **FROM** Ti **WHERE** continent = 'Asie' <br>  **UNION ALL** <br>   **SELECT** superficie, continent, pays, variete <br>   **FROM** T2 **WHERE NOT** (continent = 'Asie')) <br> **GROUP BY** continent, variete; |

Nous avons mené des expérimentations afin d'évaluer le coût de l'opération. Notre objectif est d'analyser le coût que représente cette opération calculée dynamiquement lorsque l'utilisateur en exprime le besoin. La comparaison a consisté à observer le coût théorique du calcul de requêtes portant sur ces relations de base. Nous avons comparé deux requêtes :
- une requête effectuant dynamiquement la transformation multigraduelle (série 1),
- une requête utilisant un attribut stockant le résultat de la transformation multigraduelle simulant l'utilisation d'une BDM modélisée en fonction du besoin équivalent à la transformation multigraduelle (série 2).

Nous avons généré des enregistrements dans les relations de notre BDM, complétées par des index multiples sur les clés étrangères : |ORGANISMES| = 250, 10 ≤ |GEOGRAPHIES| ≤ 100, |REPARTITION| = |ORGANISMES| x |GEOGRAPHIES|. Les valeurs des enregistrements ont été générées aléatoirement en veillant à ce que les deux sous-ensembles $E_{sup}$ et $E_{inf}$ soient de tailles homogènes.

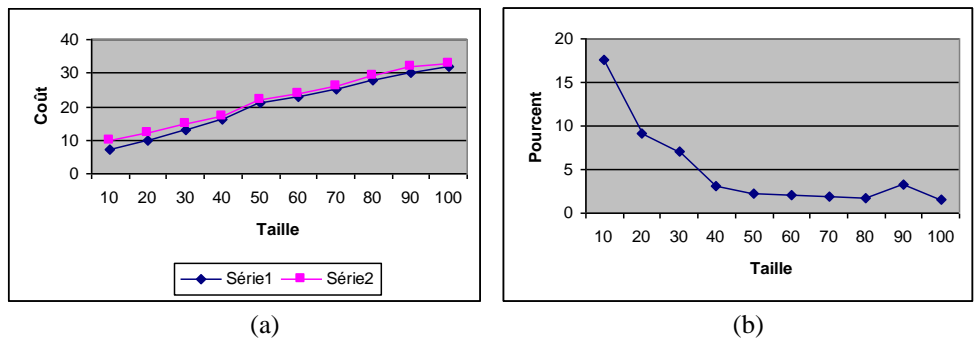

(a)  (b)

FIG. 6 – *Résultat d'expérimentation sur le coût de l'opérateur BLEND.*

Les valeurs du coût correspondent au coût théorique calculé par le SGBD (coût fourni par le « explain plan » d'Oracle Server Application 11g). En effet, notre expérimentation vise simplement à étudier le surcoût engendré par un calcul du BLEND. La taille correspond aux variations du nombre de n-uplets dans GEOGRAPHIES (10 à 100) et REPARTITION (250x10 à 250x100) ; la taille de la relation ORGANISMES reste constante à 250 n-uplets.

La figure 6(a) compare les deux requêtes. Le calcul dynamique (série 1) est évidemment plus couteux que l'utilisation de la transformation préalablement stockée (série 2), cependant, ce coût n'est pas très important (entre 18% et 2%). Ce résultat est d'autant plus encourageant que nous montrons sur la figure 6(b) que ce surcoût tend à diminuer avec l'accroissement du nombre d'enregistrements dans la relation transformée.

Néanmoins, il convient de poursuivre nos expérimentations, pour consolider ces premiers résultats. Nous envisageons plusieurs voies :
- tester notre opérateur sur une dimension plus volumineuse en augmentant significativement les tailles des relations GEOGRAPHIES et REPARTITION,
- observer si le comportement de l'opérateur reste stable avec des prédicats de calcul à la marge (lors de cette première expérimentation, nous avons utilisé un prédicat aboutissant à une partition de taille homogène entre $E_{sup}$ et $E_{inf}$) de manière à obtenir un $E_{sup}$ disproportionné par rapport à $E_{inf}$, ou encore $E_{sup}$ ou $E_{inf}$ vide,

- intégrer dans nos scénarii de tests des vues matérialisées optimisant le temps de calcul des requêtes afin de s'assurer que le coût de l'opérateur reste supportable par comparaison à un stockage pré-calculé.

## 5  Conclusion

Cet article traite d'un type d'analyses complexes consistant à combiner des paramètres de niveaux de granularité différents. Ce type d'analyses dites multigraduelles est difficilement réalisable avec les systèmes traditionnels puisqu'elles nécessitent d'organiser les données en fonction de chaque analyse. Nous introduisons un nouvel opérateur algébrique pour les manipulations OLAP, appelé « BLEND ». Nous étudions les contours de son utilisation sur des hiérarchies strictes. L'approche permet de transformer une hiérarchie en maintenant les autres possibilités de navigation sur celle-ci. L'opérateur proposé est mis en œuvre dans le contexte R-OLAP au sein du SGBD Oracle, afin d'établir la faisabilité de la proposition.

A court terme, une première perspective est de mener une étude sur les techniques d'optimisation possibles de l'opérateur, notamment en exploitant les treillis de vues matérialisées mis en place au sein des BDM. L'expression de l'opération dans notre langage graphique (Ravat, *et al.*, 2007) constitue également une extension directe de ces travaux. Nous projetons également d'étudier d'autres principes de transformations multigraduelles dans des contextes plus complexes tels que celui des hiérarchies non strictes.

## Références

## Summary

Decisional systems are based on multidimensional databases improving OLAP analyses. The paper describes a new OLAP operator named « BLEND » to perform multigradual analyses. The operation transforms multidimensional structures during querying in order to analyse measures according to various granularity levels, which are reorganised into a single parameter. We study valid combinations of the operation in the context of strict hierarchies. First experimentations implement the operation in an R-OLAP framework showing the slight cost of this operation.